\documentclass{ws-ijmpa}

\begin{document}


%
\catchline{}{}{}{}{}
%

\title{Charm Dalitz analyzes from CLEO-c 
\footnote{Presented at Charm-2006 conference, June 5-7, IHEP, Beijing, China.}
}

\author{Mikhail Dubrovin 
}

\address{Wayne State University, 666 W. Hancock St \\
Detroit, 48201 MI,
U.S.A.\\ 
dubrovin@mail.lepp.cornell.edu}

\author{CLEO Collaboration}

\address{Wilson Synchrotron Laboratory, Cornell University\\
Ithaca, NY 14853, U.S.A.}

\maketitle

\begin{history}
\today
\end{history}

\begin{abstract}
I discuss CLEO-c opportunities in Dalitz plot analyzes with
the data samples available now and projected by the end of
CESR-c run.
Using 281~pb$^{-1}$ of $e^+e^-$ collisions at mass of $\psi(3770)$
we present results of the Dalitz plot analysis
of $D^+ \to \pi^-\pi^+\pi^+$.
Using the CLEO-c and CLEO III samples of 5~pb$^{-1}$, accrued at mass of $\psi(2S)$,
we study three body decays of $\chi_{cJ}$ produced in the radiative decay
$\psi(2S) \to \gamma  \chi_{cJ}$, where $J$=0,1,2.
A clear signal from at least one of $\chi_{cJ}$ is found in eight final states:
$\pi^+\pi^-\eta$,
$\pi^+\pi^-\eta^\prime$,
$K^+K^-\pi^0$,
$K^0_S K \pi$,
$\eta K^+K^-$,
$\pi^0 p\bar{p}$,
$\eta p\bar{p}$, and
$\Lambda K^+\bar{p}$.
For these modes we measured or set an upper limit
on the branching fraction.
A Dalitz plot analysis is performed on three modes 
$\chi_{c1} \to \pi^+\pi^-\eta$, $K^+K^-\pi^0$, and $K^0_S K \pi$.

\keywords{charmonium; meson; production; decay; Dalitz; CLEO; experiment.}
\end{abstract}

\ccode{PACS numbers: 13.25.Ft, 13.25.Gv, 13.25.Jx, 13.60.Le, 13.66.Bc}

\section{The CLEO-c opportunities in Dalitz plot analyzes}	

Currently the CLEO collaboration possesses multiple samples of data
which can be used for analysis  
of $D$, $D_s$, $\chi_{cJ}$, $J/\psi$ meson decays. 
\\$\bullet$
     A sample of 
     $e^+e^- \to \psi(3770) \to D\overline{D}$ 
     events with integrated luminosity of 281~pb$^{-1}$, 
     which is equivalent to 
     $\sim10^6$ of $D^+D^-$, and $0.8\cdot10^6$ of $D^0\overline{D^0}$ pairs.
     We find about 20 three-body decay modes 
     of charged and neutral $D$ mesons 
     to the combinations of $\pi$, $K$, and $\eta$ in the final state
     that can be used for the Dalitz plot analysis.
     Running at $D\overline{D}$ production threshold gives many advantages. 
     There is no extra-energy for
     production of additional fragmentation particles which keeps the multiplicity low
     and gives very clean events.
     In some cases, where combinatoric background is small, 
     we use inclusive reconstruction of a single D meson.
     In exclusive modes one $D$ meson of $D\overline{D}$ pair can be used as a flavor or CP tag \cite{CPtagging},
     and the recoiling $D$ studied.
\\$\bullet$
     A sample of 
     $e^+e^- \to D_s^*\overline{D_s}$ events at $\sqrt{s}=4170$~MeV with
     ${\mathcal L}\simeq$200~pb$^{-1}$,
     or $2\cdot10^5$ of $D_s^*\overline{D_s}$ (charge conjugated modes are implied throughout this paper).
     This sample has been collected during Spring of this year
     and its processing is still in progress.
     We are going to use $D_s$ mesons for the Dalitz plot analysis.
     In particular, for the decay mode $D_s^+ \to \phi\pi^+$, often used as a reference,  
     an interference with other modes can be studied in the 
     amplitude analysis of the $D_s^+ \to K^+K^-\pi^+$ decay.
\\$\bullet$
     A sample of 
     $e^+e^- \to \psi(2S)$ events with 
     ${\mathcal L}\simeq$5~pb$^{-1}$, 
     or 3 million produced $\psi(2S)$'s which can be used to study
     $J/\psi$ or $\chi_{cJ}$ decays in the processes
     \begin{itemize}
     \item[$\rhd$]
           $\psi(2S) \to \pi \pi J/\psi$  with ${\mathcal B}\simeq 50\%$
           for $\pi^+ \pi^-$ and  $\pi^0 \pi^0$ together,
     \item[$\rhd$]
           $\psi(2S) \to \gamma \chi_{cJ}$  with ${\mathcal B}\simeq 9\%$ for each spin $J=0,1,2$ state.
     \end{itemize}
     The $\pi \pi$ pair in the first case and the $\gamma$ in the second can be used
     as an additional signal tag for these processes. 
     A partial wave analysis can be applied in order to study decay of particle with non-zero spin,
     but as a first step, with this small sample in hand, we perform Dalitz plot analyzes with
     a special treatment of angular distributions.  

The current CLEO-c run plan
has running on the $\psi(2S)$ completed during the fall of 2006
accumulating ten times more statistics of $\psi(2S)$.
By the end of the CLEO-c program (April 2008) we are going to gather
samples of $D\overline{D}$  three and
$D_s^*\overline{D_s}$ four times larger than we have at present. 
This data size and cleanliness is encouraging us to proceed with decay substructure study 
in 3-body decay modes using the Dalitz plot technique~\cite{Dalitz}.  
At this conference we present preliminary results of a few
Dalitz plot analyzes of the decays $D^+ \to \pi^-\pi^+\pi^+$ and
$\chi_{c1} \to \pi^+\pi^-\eta$, $K^+K^-\pi^0$, and $K^0_SK\pi$.


\section{Dalitz plot analysis of  $D^+ \to \pi^-\pi^+\pi^+$}

A Dalitz plot analysis of $D^+ \to \pi^-\pi^+\pi^+$ has previously been done by
E791~\cite{791} and 
FOCUS~\cite{focus}.  The preliminary analysis described here is from
CLEO-c~\cite{CLEOdet}, and represents the first time we have done the 
same Dalitz plot analysis as the fixed target experiments.
The decay is selected with cuts on the beam constrained mass
of three charged tracks consistent with pions, 
$m_{BC}=\sqrt{E^2_{beam}-p^2(\pi^-\pi^+\pi^+)}$, 
and the difference of their energy
from the beam energy,
$\Delta E = E(\pi^-\pi^+\pi^+) - E_{beam}$. 
A sample of 6991 events is selected with a signal to noise of about two to one. 
The E791 and FOCUS samples are of similar size and cleanliness.

\begin{figure*}[t]
\includegraphics[width=40mm]{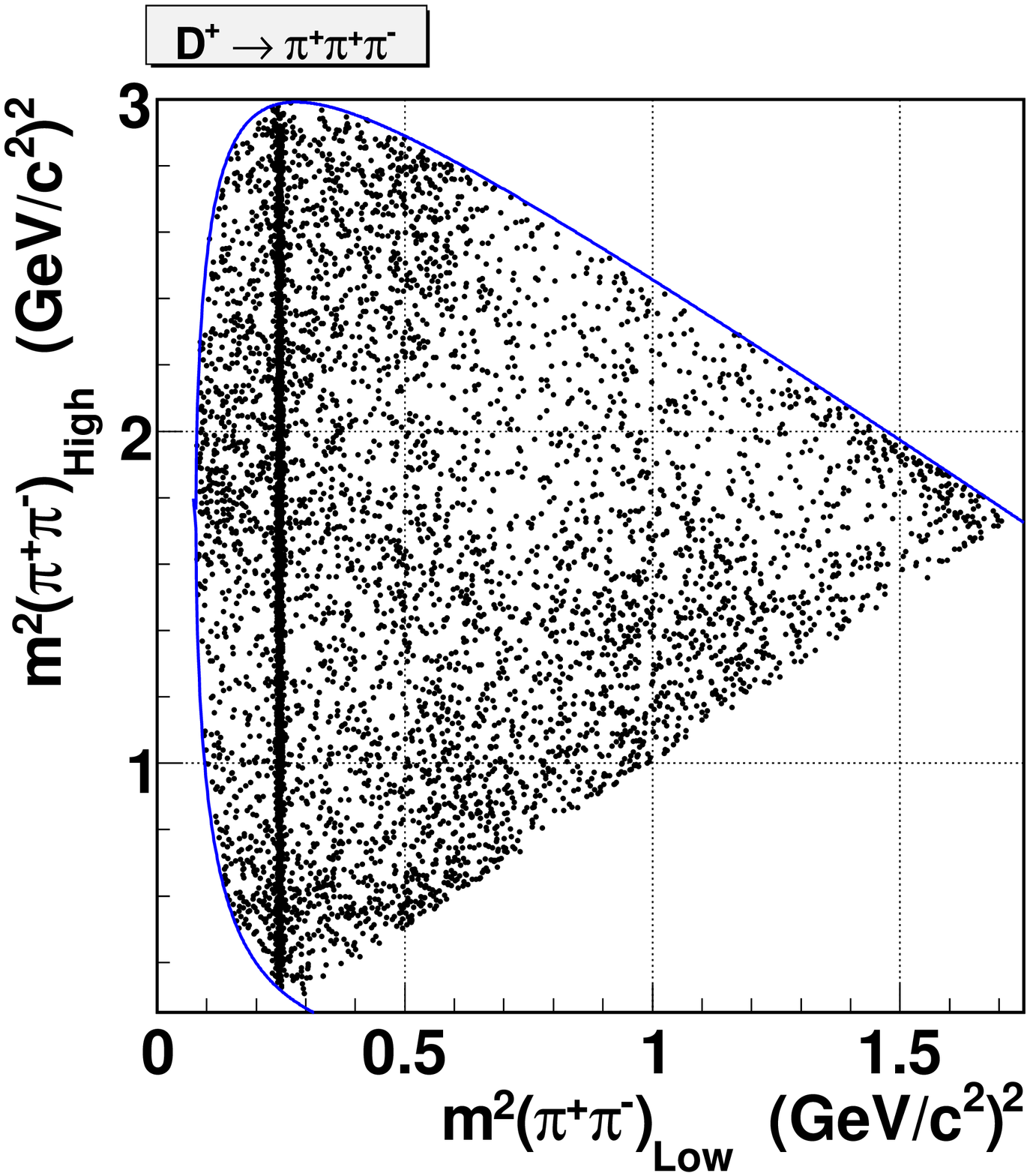}
\includegraphics[width=40mm]{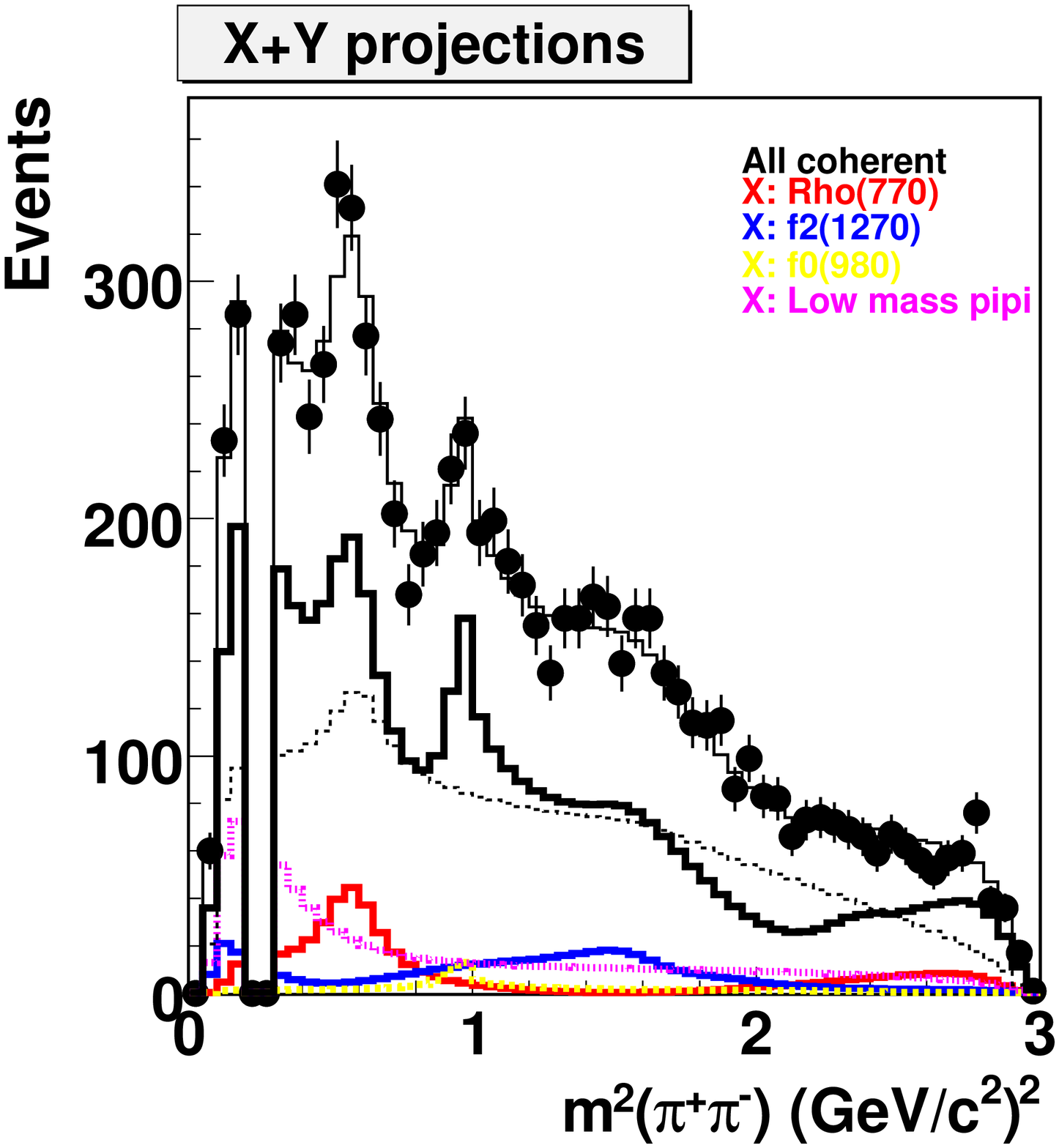}
\includegraphics[width=40mm]{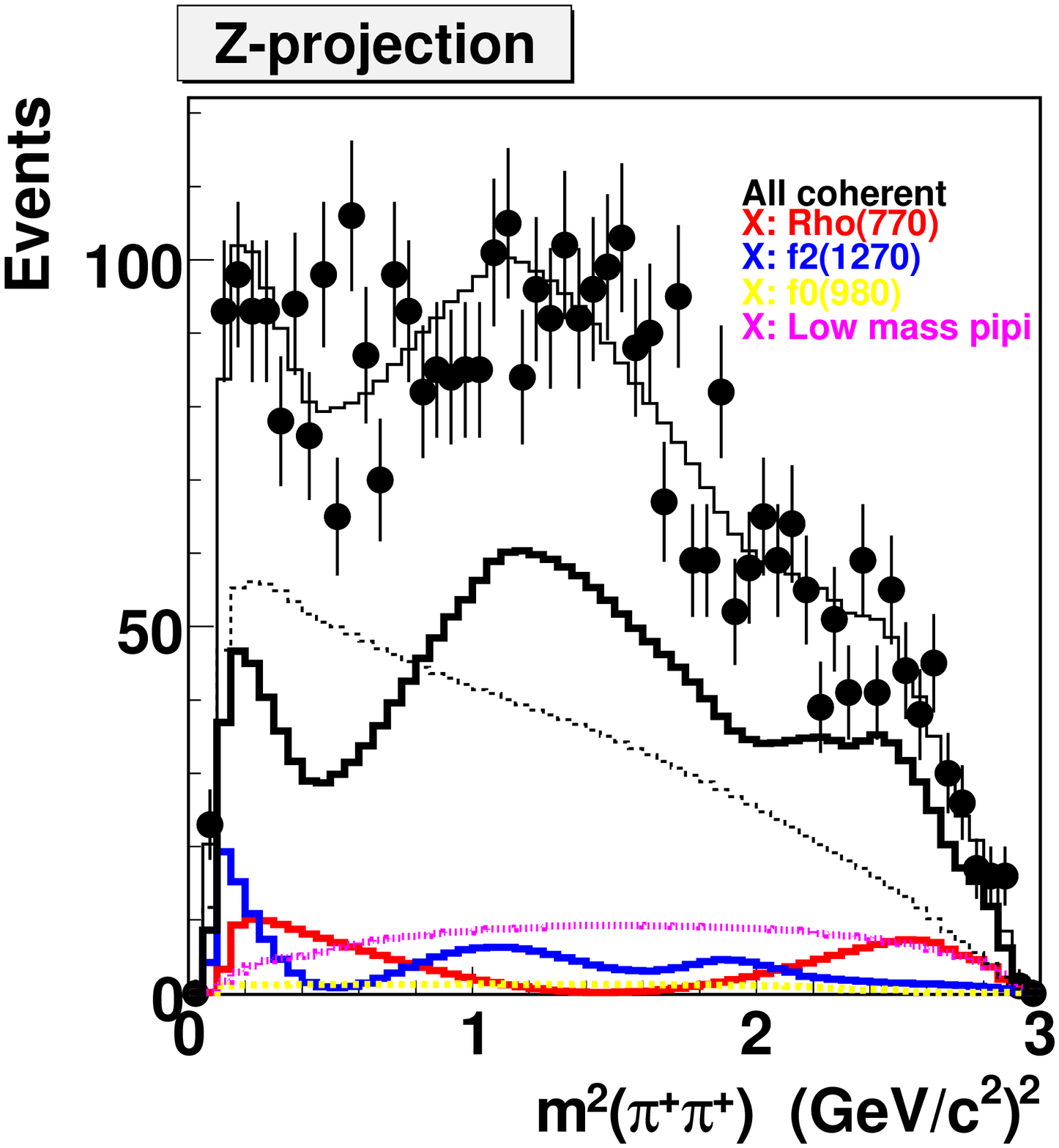}
\caption{Dalitz plot and projections for $D^+ \to \pi^-\pi^+\pi^+$.} 
\label{fig:d3pidalitz}
\end{figure*}

The Dalitz plot is symmetric under the interchange of like-sign pions thus we
do the fit in the two dimensions of high versus low unlike-sign pion mass,
as shown in Fig.~\ref{fig:d3pidalitz}(left).  
There is a large contribution from $D^+ \to K^0_{\rm S}\pi^+$ which
because of the long $K^0_{\rm S}$ lifetime should not interfere with other submodes.  
We do not consider events with $m(\pi^+\pi^-)$ within ten standard
deviations in mass resolution of the known $K^0_{\rm S}$ mass on the Dalitz plot.  This
leaves 4068 events which are 
shown in the projection plots of Fig.~\ref{fig:d3pidalitz}. 
An unbinned maximum likelihood fit is used along with other methods
as described in Ref.~\cite{cleodalitz}. 
The efficiency across the Dalitz plot is modeled with simulated events that are
fit to a two-dimensional second order polynomial and threshold functions to account
the fall off at the corners.  
Backgrounds are taken from $m_{BC}$ and $\Delta E$ sidebands and 
extra resonance contributions are allowed from mis-measured $K^0_{\rm S}$, $\rho$,
and $f_2(1270)$ decays.  
A total of 13 different resonances are considered.  
Parameters describing these
resonances are taken from previous experiments.  Only contributions with an
amplitude significant at more than three standard deviations are said to
be observed, and others are limited.  Contributions that are not significant are
not included in the decay model used for the result.

Contributions from $\rho^0\pi$, $f_0(980)\pi$, and $f_2(1270)\pi$ are clearly visible.  
Table~\ref{tab:d3pidalitz} shows the preliminary amplitudes, phases, 
and fit fractions measured by CLEO
compared with the results of the E791 analysis mentioned above.  
\begin{table}[ph]
\tbl{ Summary of results for $D^+ \to \pi^-\pi^+\pi^+$.}
{\begin{tabular}{ c|c|c|c||c }
Mode             & Amplitude           & Phase ($^\circ$)  & Fit Fraction (\%)    & E791: Fit Fraction (\%) \\
\hline		  
$\rho(770)\pi^+$ & 1(fixed)            & 0(fixed)          & 20.0$\pm$2.3$\pm$0.9 & 33.6$\pm$3.9 \\
$f_0(980)\pi^+$  & 1.4$\pm$0.2$\pm$0.2 &  12$\pm$10$\pm$5  & 4.1$\pm$0.9$\pm$0.3  & 6.2$\pm$1.4  \\
$f_2(1270)\pi^+$ & 2.1$\pm$0.2$\pm$0.1 &-123$\pm$6$\pm$3   & 18.2$\pm$2.6$\pm$0.7 & 19.4$\pm$2.5 \\
$f_0(1370)\pi^+$ & 1.3$\pm$0.4$\pm$0.2 & -21$\pm$15$\pm$14 & 2.6$\pm$1.8$\pm$0.6  & 2.3$\pm$1.7  \\
$f_0(1500)\pi^+$ & 1.1$\pm$0.3$\pm$0.2 & -44$\pm$13$\pm$16 & 3.4$\pm$1.0$\pm$0.8  & ---  \\
$\sigma\pi^+$    & 3.7$\pm$0.3$\pm$0.2 &  -3$\pm$4$\pm$2   & 41.8$\pm$1.4$\pm$2.5 & 46.3$\pm$9.2 \\
\hline		  								                         
Non-resonant     &                     &                   & $<$3.5               & 7.8$\pm$6.6  \\
$\rho(1450)\pi^+$&                     &                   & $<$2.4               & 0.7$\pm$0.8  \\
\hline		  								                         
$\sum_i FF_i$, \%&  \multicolumn{2}{|c|}{  }               &  90.2                & 116          \\
\end{tabular} \label{tab:d3pidalitz} }
\end{table}
There is broad agreement between the two results, including the observation of a $\sigma\pi$
contribution. In an alternative fit using the same decay model as E791
the agreement is only slightly better, but the fit is much less likely
to model our data than the model shown in the table which does not include
non-resonant and $\rho(1450) \pi^+$ contributions, but does include a $f_0(1500) \pi^+$
contribution.  Models without a $\sigma\pi$ contribution do not agree well 
with the data.

This CLEO analysis is preliminary, and we plan to consider a generalized
model of $\pi^+\pi^-$ S-wave interactions to model $\sigma$ and $f_0$ contributions
such as the K-matrix which is used in the FOCUS analysis mentioned above.


\section{Three-body decays of $\chi_{cJ}$}

Decays of the $\chi_{cJ}$ states are not well studied both
experimentally and theoretically.
Assuming that $\chi_{cJ}$ are the $^3P_J$ $c\bar{c}$-bound states
one would expect that $\chi_{c0}$ and $\chi_{c2}$ with $J^{PC}$ quantum numbers
$0^{++}$ and $2^{++}$ decay to the light quarks via two gluons~\cite{quarkoniumreview}.
Measurement of any possible $\chi_{cJ}$ hadronic decays provides
valuable information on possible glueball dynamics.

We study the decay modes of $\chi_{cJ}$ using the process
$e^+e^- \to \psi(2S) \to \gamma \chi_{cJ}$
with 3~million $\psi(2S)$ produced.
We search for signals in the $\chi_{cJ} \to h^0 h^+ h^-$ three-body decays, where
$h^0$ stands for  $\pi^0$, $K^0_s$, $\eta$, $\eta^{\prime}$ or $\Lambda$, and
$h^+$ is $\pi^+$, $K^+$ or $p$.

Our basic technique is an exclusive analysis.  A photon
candidate is combined with three hadrons and the 4-momentum sum constrained
to the known beam energy and small momentum caused by the
beam crossing angle.
We cut on the $\chi^2$ of this fit, which has four degrees of freedom,
as it strongly discriminates between background and signal.  
Efficiencies
and backgrounds are studied in a GEANT-based simulation of
the detector's full response.
Our simulated sample is roughly ten times our data sample.
The simulation is generated with a $ 1 + \lambda \cos^2\theta$
distribution in $\cos\theta$, where $\theta$ is the radiated photon angle 
relative to the positron beam axis. 
A E1 transition, as expected for $\psi(2S) \to \gamma\chi_{cJ}$,
implies $\lambda=1,-1/3,+1/13$ for $J=0,1,2$ particles. 
The differences of efficiencies 
due to various $\theta$ distributions are negligible
as we accept transition photons down to our detection limit.

We use standard CLEO algorithms for tracks reconstruction in the tracking system
and photons in $CsI$ calorimeter.
Combined $dE/dx$ and RICH, $E/p$ are exploited for particle identification
and background suppression.
$\pi^0$'s are formed from two photon candidates of the $\pi^0 \to \gamma\gamma$ decay.
We use $\eta \to \gamma\gamma$ and $\eta \to \pi^+\pi^-\pi^0$ decays to
reconstruct $\eta$ mesons.
The  $\eta^{\prime}$ candidates are formed from $\eta^{\prime} \to \eta \pi^+\pi^-$ and 
and $\eta^{\prime} \to \gamma \rho$ 
(with $E_\gamma > 200$~MeV, and $\pi^+\pi^-$  mass to be within 100~MeV/c$^2$ of the $\rho$ mass)
combinations. 
In most cases mass constrained fits are applied with appropriate cut on $\chi^2$ value.
$K^0_S \to \pi^+\pi^-$ and $\Lambda \to p\pi^-$ candidates
are formed from good quality tracks that are constrained to come
from a common vertex.
The $K^0_S$ flight path is required to be greater than 5~mm and the 
$\Lambda$ flight path greater than 3~mm.  The mass cut around the $K^0_S$ mass is 
$\pm 10$~MeV/c$^2$, and around the $\Lambda$ mass $\pm 5$~MeV/c$^2$.
Events with only the correct number of selected tracks for the
decay mode being considered are accepted.

The efficiencies averaged over the CLEO~III and CLEO-c
data sets for each mode, including the branching fractions of
$\eta \to \gamma\gamma$, 
$\eta \to \pi^+\pi^-\pi^0$ and 
$\eta^\prime \to \eta\pi^+\pi^-$,
are in the range 8\% to 30\%.

\begin{figure}
\includegraphics*[width=41mm]{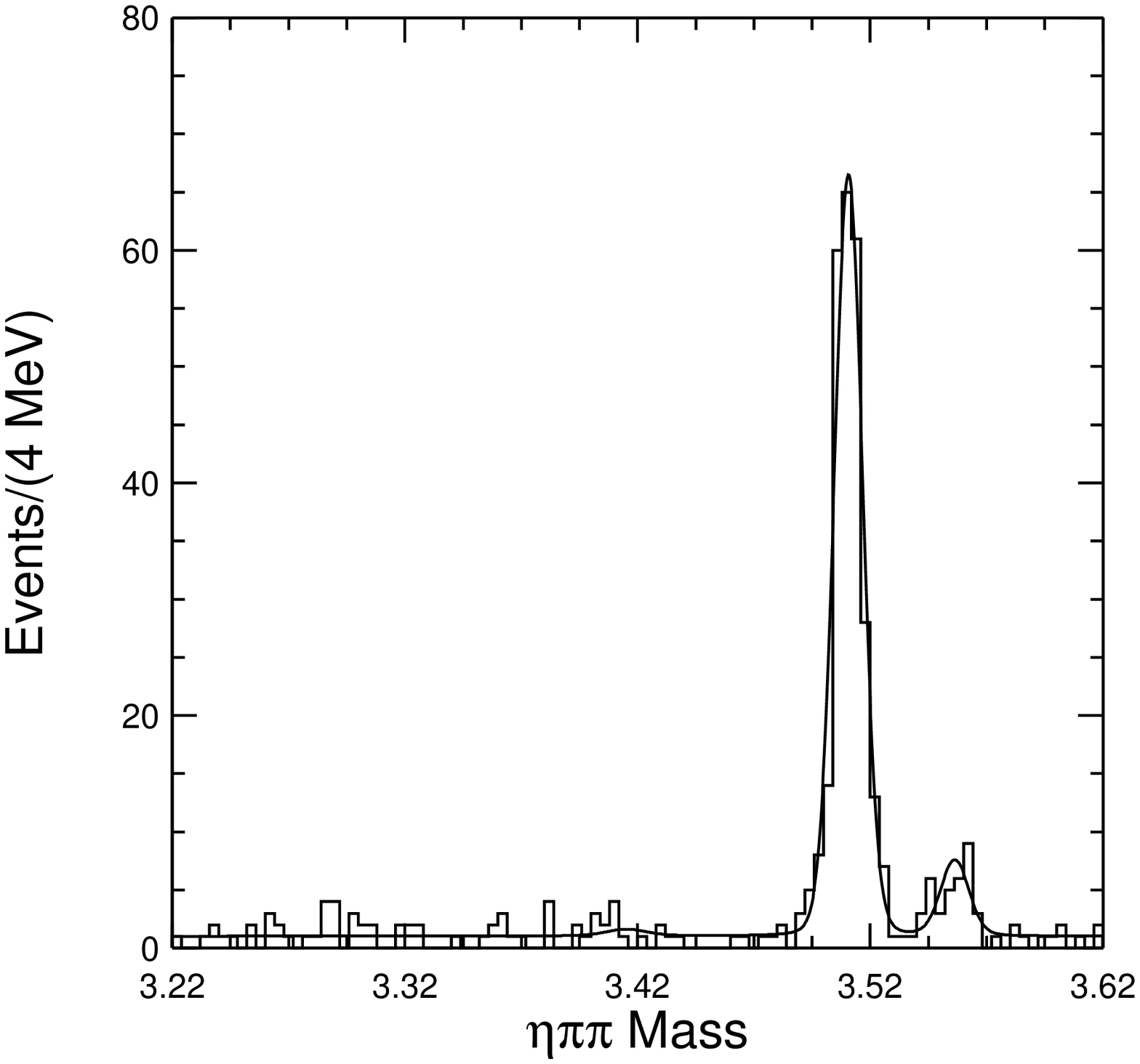}
\includegraphics*[width=41mm]{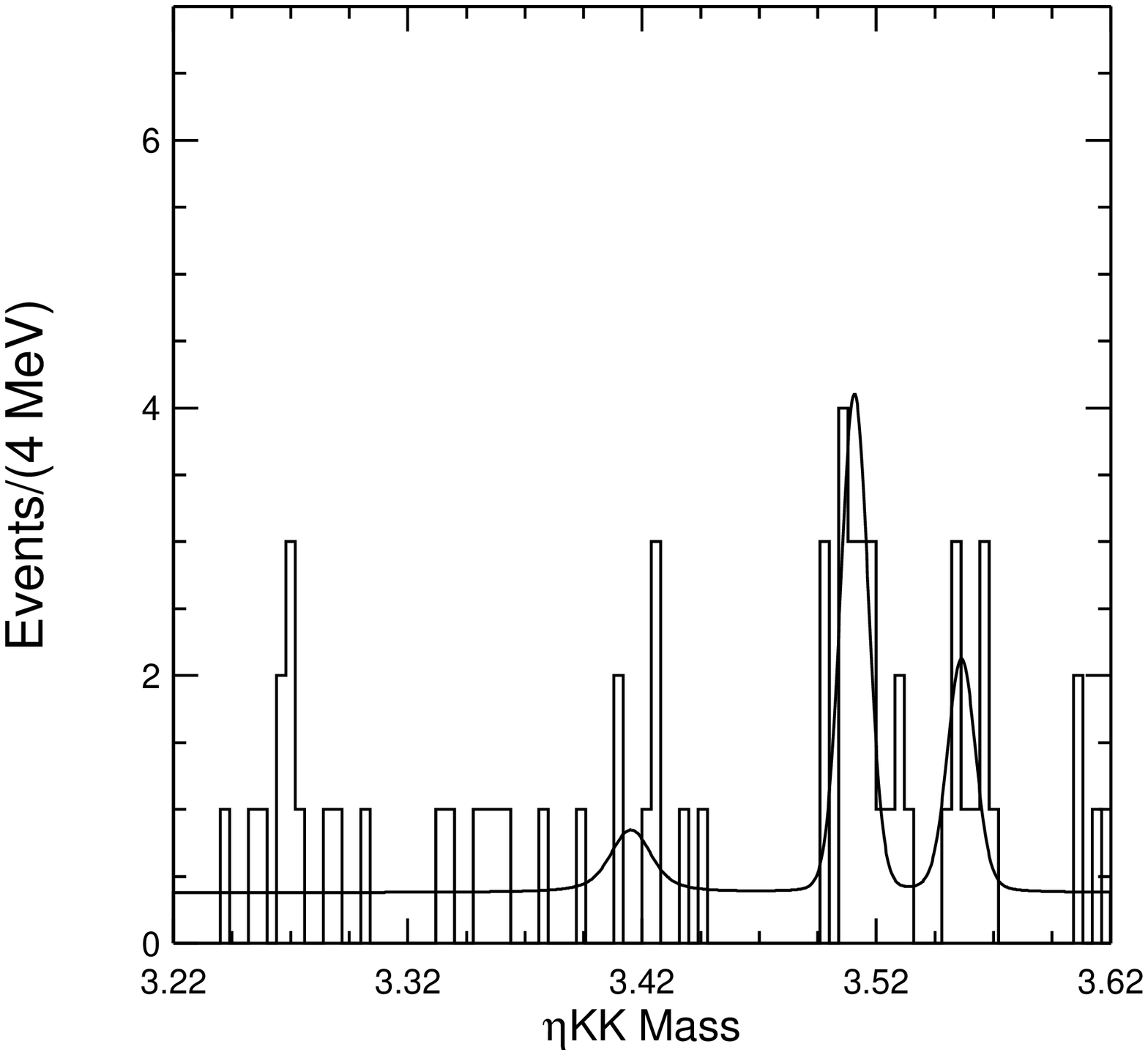}
\includegraphics*[width=41mm]{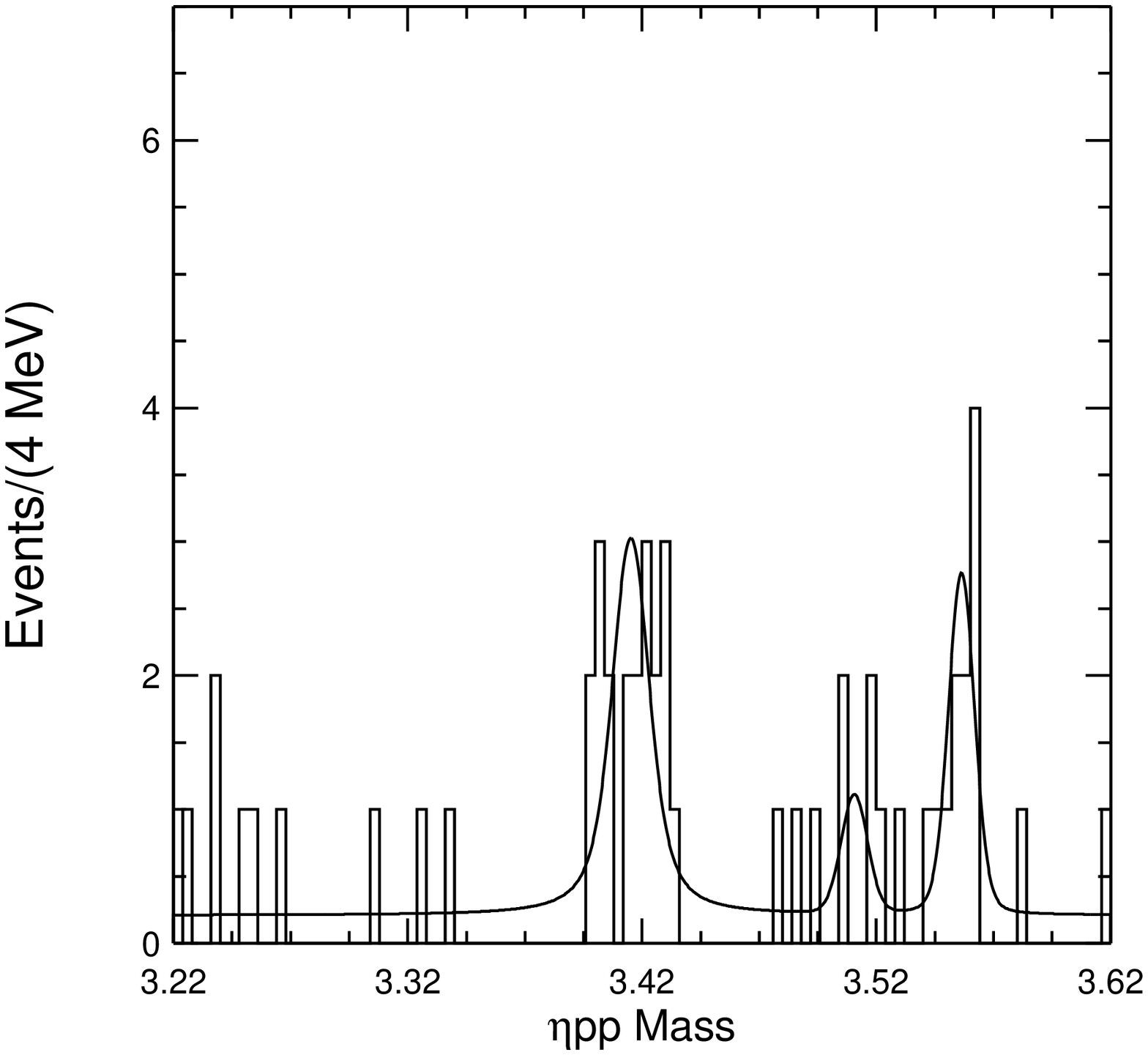}
\includegraphics*[width=41mm]{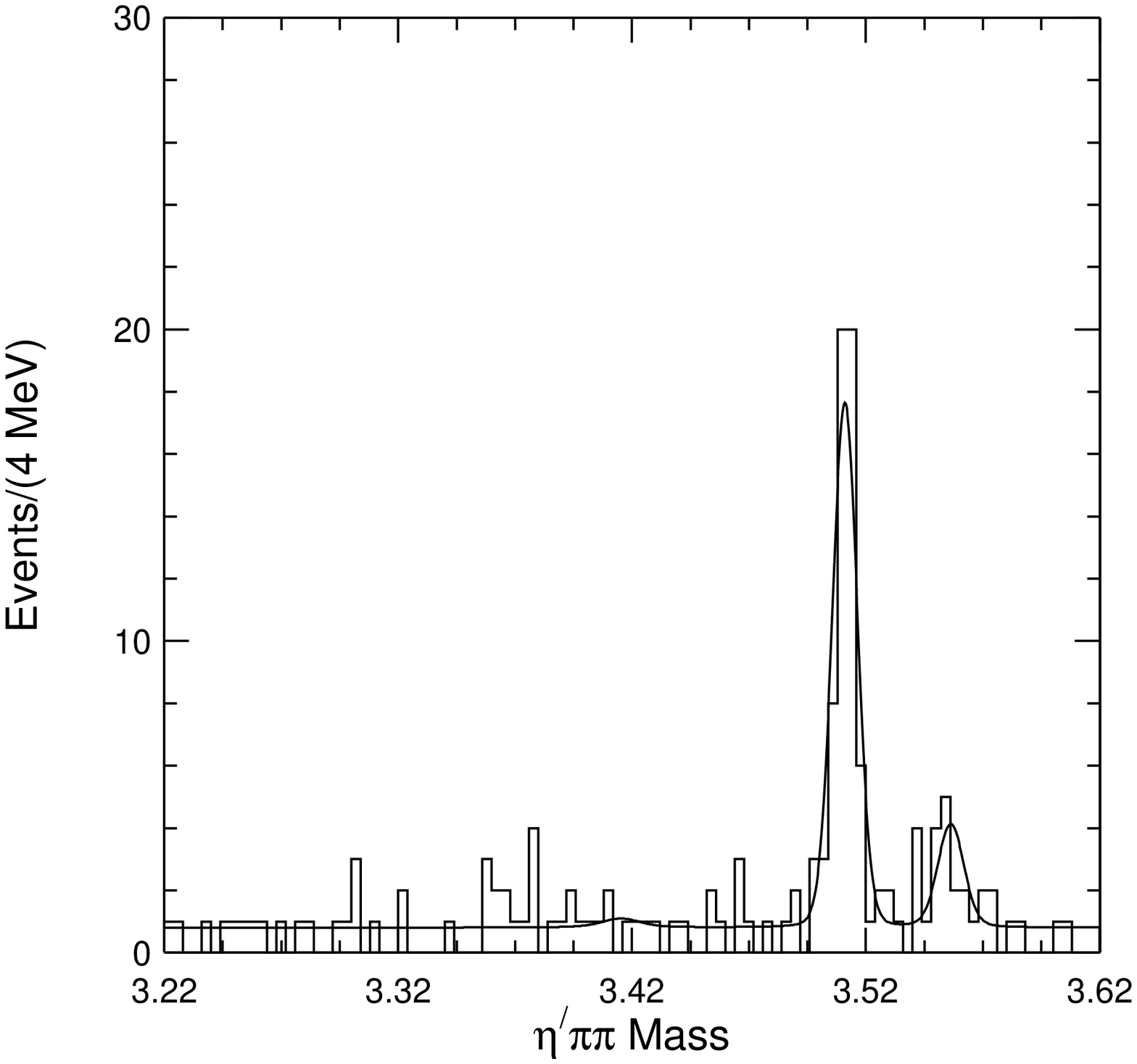}
\includegraphics*[width=41mm]{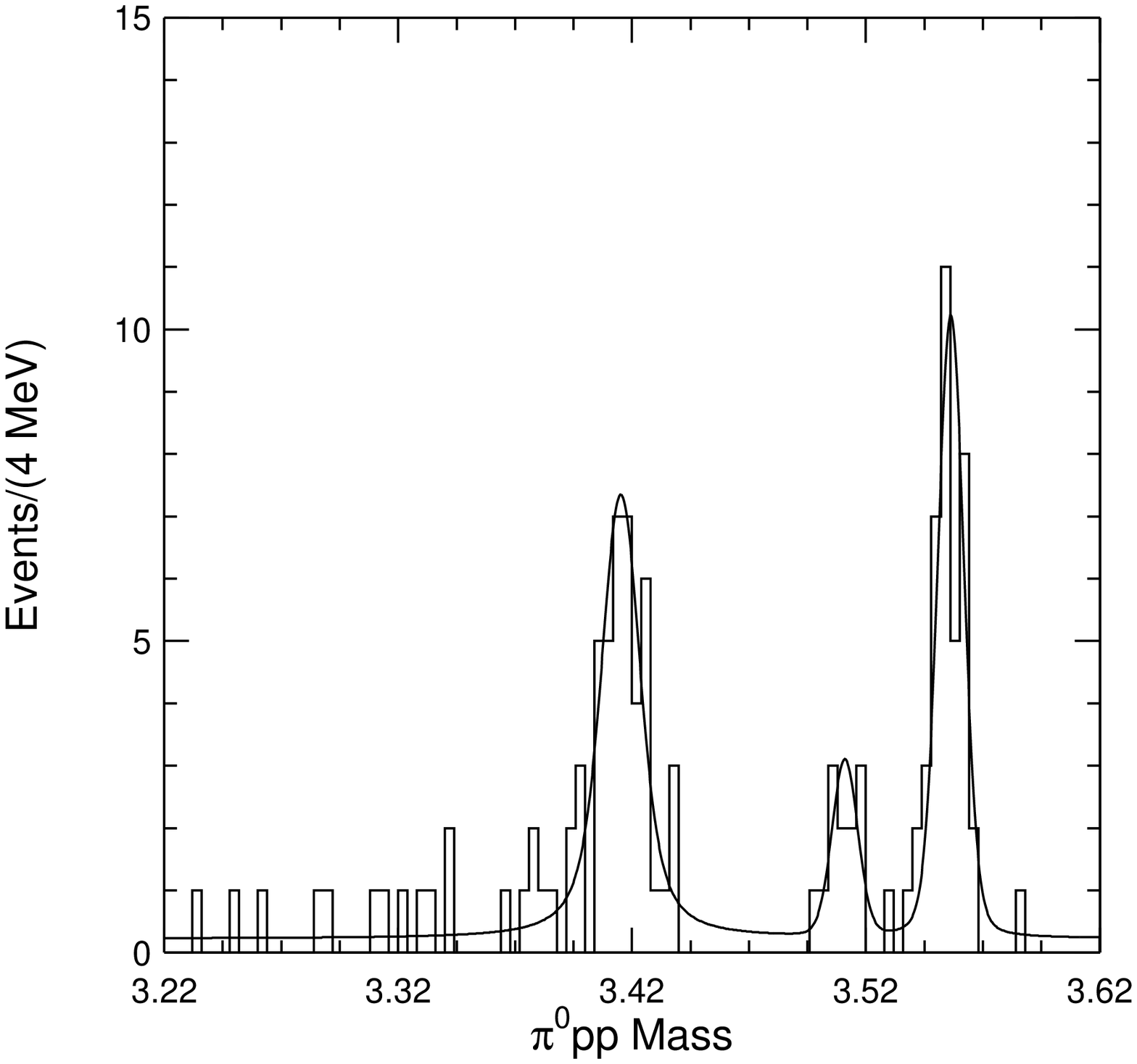}
\includegraphics*[width=41mm]{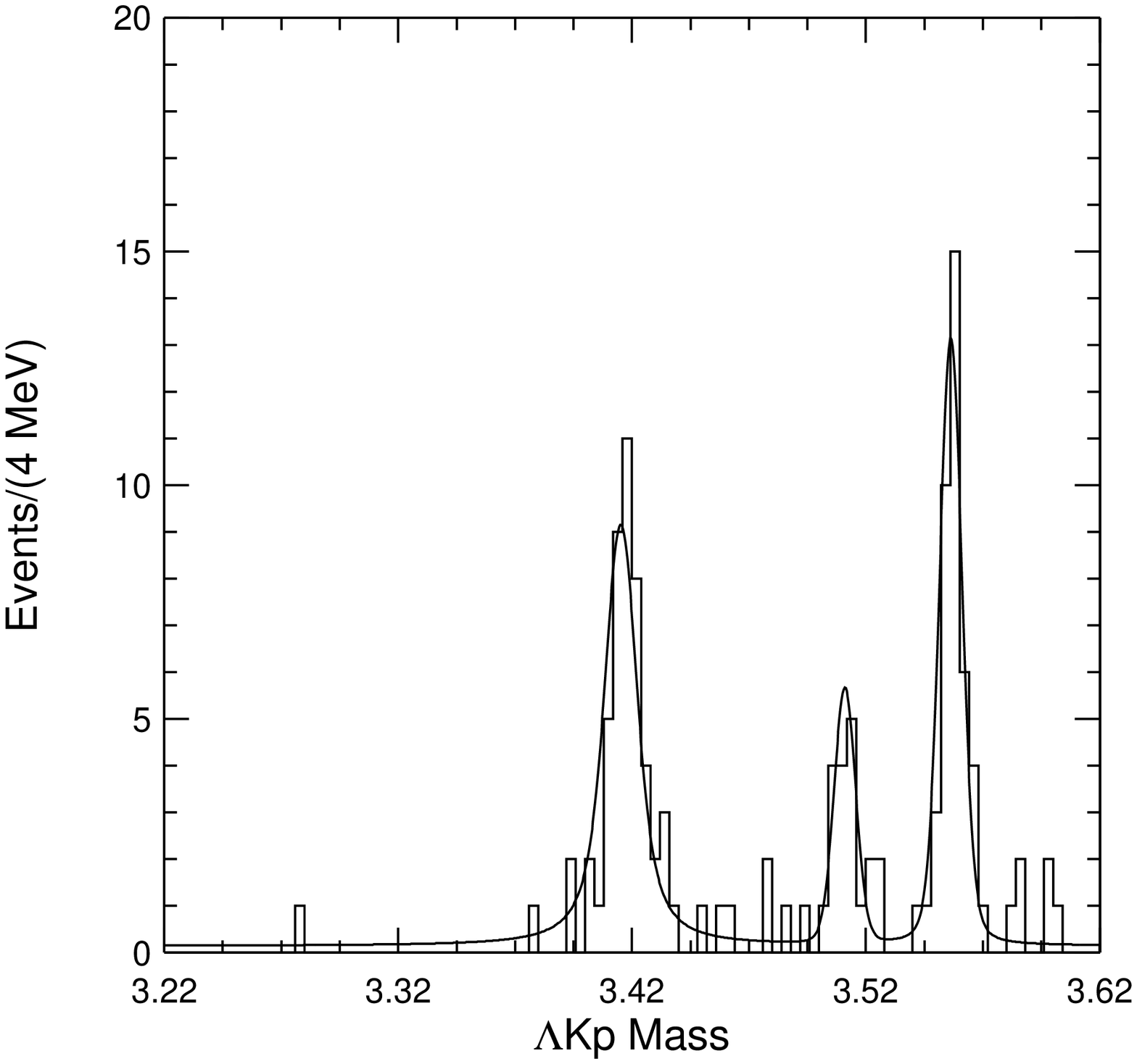}
\includegraphics*[width=41mm]{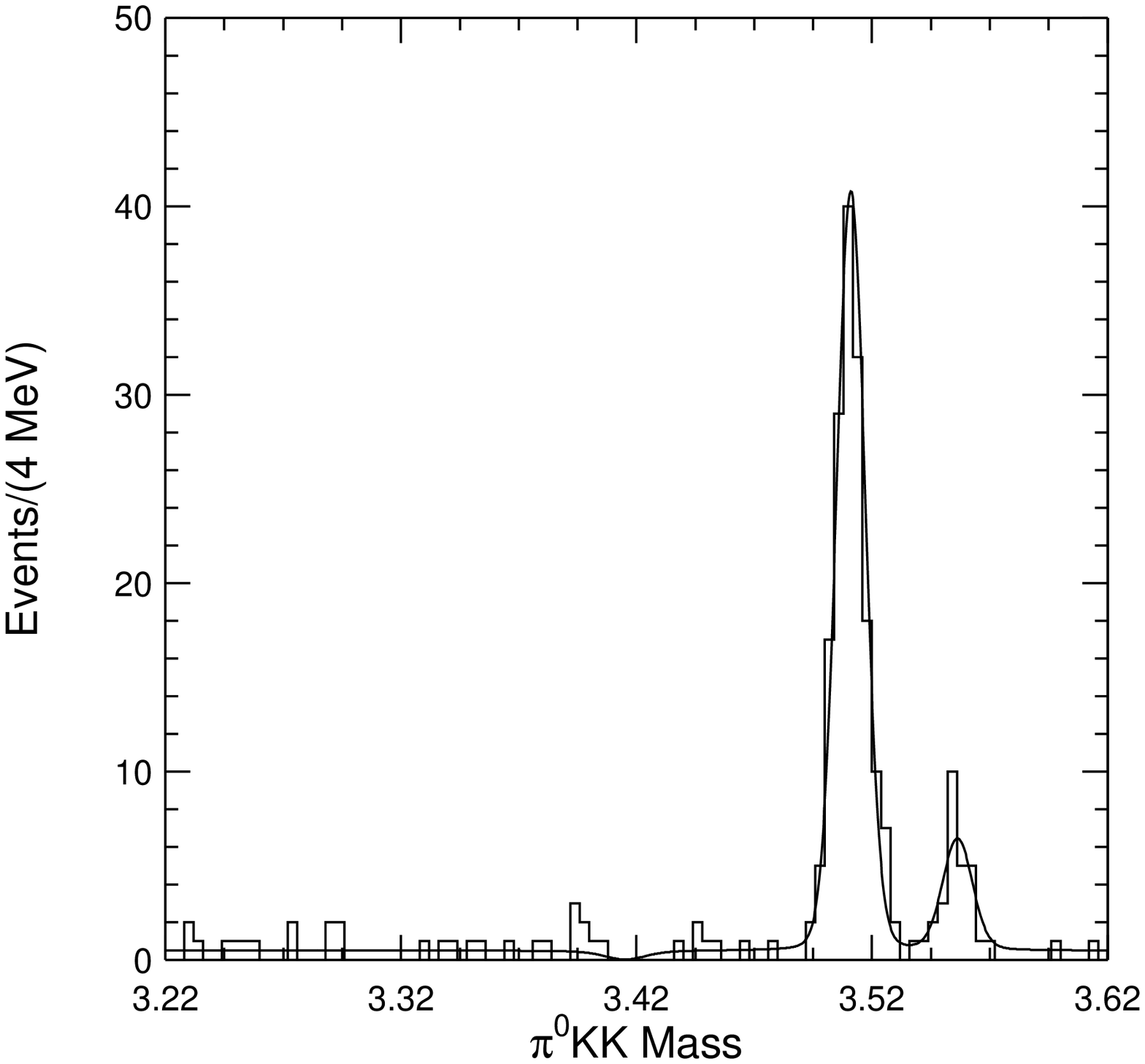} 
\includegraphics*[width=41mm]{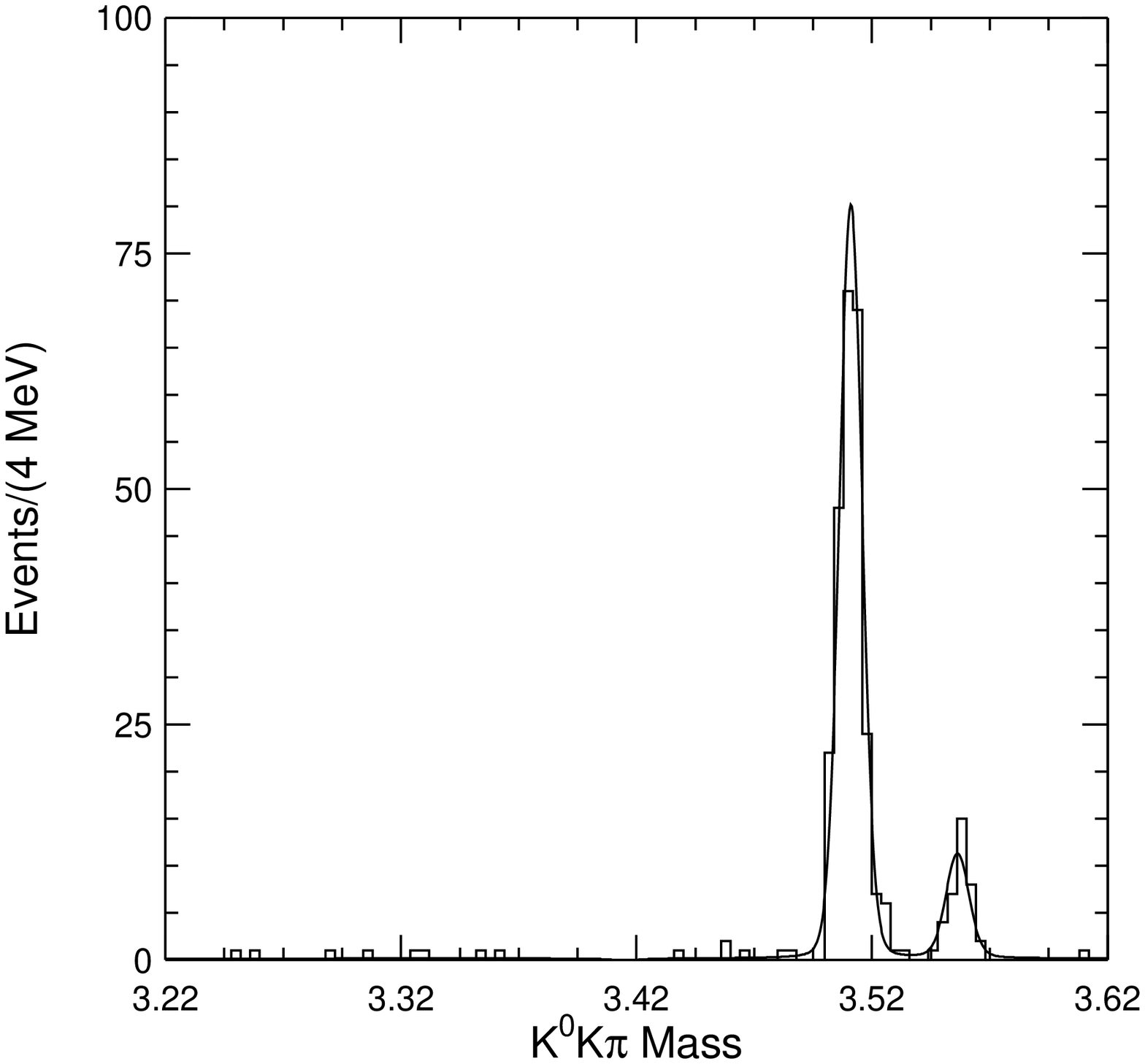}
\caption{Mass distribution for candidate 
         $\chi_{cJ} \to \pi^+\pi^-\eta$, 
         $K^+K^-\eta$, 
         $p{\bar p}\eta$, 
         $\pi^+\pi^-\eta^\prime$,
         $p {\bar p} \pi^0$, 
         $K^+{\bar p}\Lambda$,
         $K^+K^-\pi^0$, and
         $K^0_{\rm S}K^-\pi^+$.
        }
\label{fig:pipieta}
\end{figure}
Figure \ref{fig:pipieta} shows the mass
distributions with fits for the eight $\chi_{cJ}$ decay modes.
Signals are evident in all three $\chi_{cJ}$ states, but not in all the modes.  
Backgrounds are small.  
The mass distributions are fit to three signal shapes,
Breit-Wigners convolved with Gaussian detector resolutions
(with $\sigma$ from 4.3 to 7~MeV/c$^2$, depending on mode), and a linear
background.  The $\chi_{cJ}$ masses and intrinsic 
widths are fixed at the values from Ref.~\cite{pdg}.  

We consider various sources of systematic uncertainties on the yields:
allowing the $\chi_{cJ}$ masses and intrinsic widths to float in the fit; 
change the shape of background; 
break up the sample into CLEO~III and CLEO-c data sets;
uncertainty on the efficiency of photon and track reconstruction;
uncertainty due to the cut on the $\chi^2$ in mass and event vertex
constrained fits. 
Based on the results of the Dalitz plot analyzes
we correct the efficiency in the $\chi_{c1} \to \pi^+ \pi^-\eta $, $\chi_{c1} \to K^+K^-\pi^0$, and
$\chi_{c1} \to K^0_SK^-\pi^+$ modes by a relative up to $6$\% to account for the change
in the efficiency caused by the deviation from a uniform phase space distribution
of decay products to what we actually observe.
To calculate $\chi_{cJ}$ branching fractions, we use previous CLEO measurements~\cite{chicbf}
for the $\psi(2S)\to\gamma\chi_{cJ}$ branching fractions of 
$9.33\pm.14\pm.61$\%, 
$9.07\pm.11\pm.54$\% and 
$9.22\pm.11\pm.46$\%
for $J=0,1,2$ respectively.
Preliminary results for the three body branching fractions are shown in 
Table~\ref{tab:Branching_fractions}.  Where the yields do not show clear signals
we calculate 90\% confidence level upper limits accounting for statistical and
systematic uncertainties.

\begin{table}[ph]
\tbl{Preliminary branching fractions in \%. Uncertainties are statistical,
     systematic,
     and a separate systematic due to uncertainties in the $\psi(2S)$ branching
     fractions.  Limits are at the 90\% confidence level.}
{\begin{tabular}{l|c|c|c}
Mode                      & $\chi_{c0}$
                          & $\chi_{c1}$
                          & $\chi_{c2}$ \\
\hline
$\pi^+\pi^-\eta $         & $<0.021$
                          & $0.52\pm.03\pm.03\pm.03$
                          & $0.051\pm.011\pm.004\pm.003$ \\
$K^+K^-\eta$              & $<0.024$
                          & $0.034\pm.010\pm.003\pm.002$
                          & $<0.033$ \\
$p\bar{p}\eta$            & $0.038\pm.010\pm.003\pm.02$
                          & $<0.015$
                          & $.019\pm.007\pm.002\pm.002$ \\
$\pi^+\pi^-\eta^{\prime}$ & $<0.038$
                          & $0.24\pm.03\pm.02\pm.02$
                          & $<0.053$     \\
$K^+K^-\pi^0$             & $<0.006$
                          & $0.200\pm.015\pm.018\pm.014$
                          & $0.032\pm.007\pm.002\pm.002$ \\
${\bar K^0}K^-\pi^+$      & $<0.010$
                          & $0.84\pm.05\pm.06\pm.05$
                          & $0.15\pm.02\pm.01\pm.01$ \\
$p\bar{p}\pi^0$           & $0.059\pm.010\pm.006\pm.004$
                          & $0.014\pm.005\pm.001\pm.001$
                          & $0.045\pm.007\pm.004\pm.003$ \\
$K^+ \bar{p}\Lambda$      & $0.114\pm.016\pm.009\pm.007$
                          & $0.034\pm.009\pm.003\pm.002$
                          & $0.088\pm.014\pm.07\pm.006$
\end{tabular}\label{tab:Branching_fractions}}
\end{table}
%



\section{Dalitz plot analysis of $\chi_{c1} \to \pi^+\pi^-\eta$, $K^+K^-\pi^0$, and $K^0_S K \pi$}

We choose the three high statistics signals 
$\chi_{c1} \to \pi^+ \pi^-\eta $, $\chi_{c1} \to K^+K^-\pi^0$, and
$\chi_{c1} \to K^0_SK^-\pi^+$ 
with 228, 137, and  234 events respectively for Dalitz plot analyzes 
to study resonance substructure. 
Events were selected within 10 MeV, roughly two standard deviations, 
of the signal peak.
When fitting the data Dalitz plot the small contributions from backgrounds
are neglected.  We use an isobar model to describe resonance contributions
taking into account spin and width dependent effects. 
Narrow resonances are described with a Breit-Wigner amplitude with the resonance
parameters taken from previous experiments \cite{pdg}.
We use the Flatt\'e line-shape for $a_0(980)$ with parameters from the
Crystal Barrel Collaboration~\cite{CBarrel_a0_980}. 
For low mass $\pi^+\pi^-$($\sigma$) and  $K\pi$($\kappa$) S-wave contributions
we choose a simple description which is adequate for our small sample~\cite{Oller_2005}.

We are examining the $e^+e^- \to \psi(2S) \to \gamma \chi_{c1}$ process.
In such a decay the $\chi_{c1}$ should be polarized.  
For our small sample we use angular distributions 
from~\cite{Filippini-Fontana-Rotondi}, calculated for non-polarized $\chi_{c1}$ decay.
We have tested different angular distributions and 
include the variations as a systematic uncertainty. 

\begin{figure}
\includegraphics*[width=41mm]{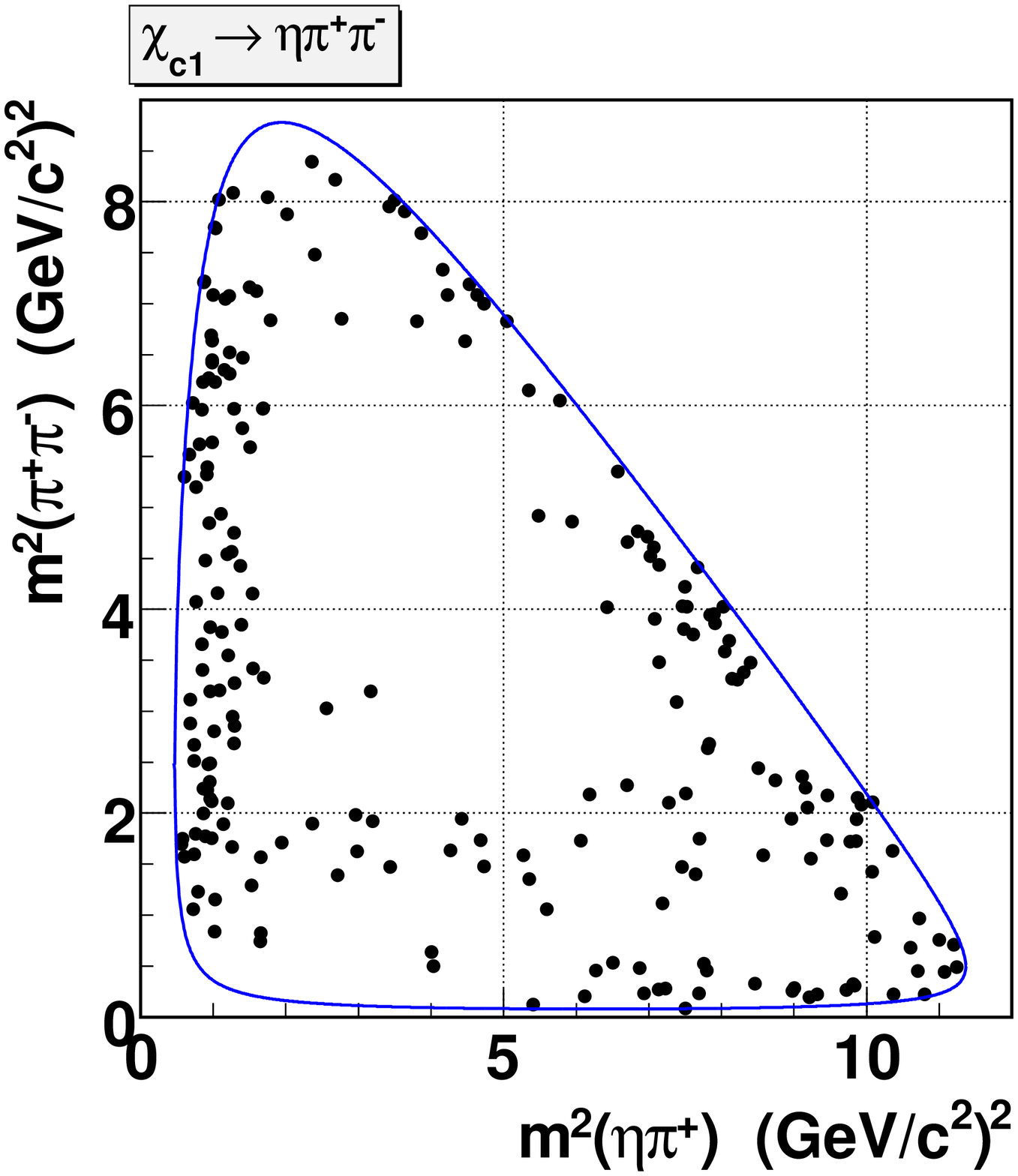}
\includegraphics*[width=41mm]{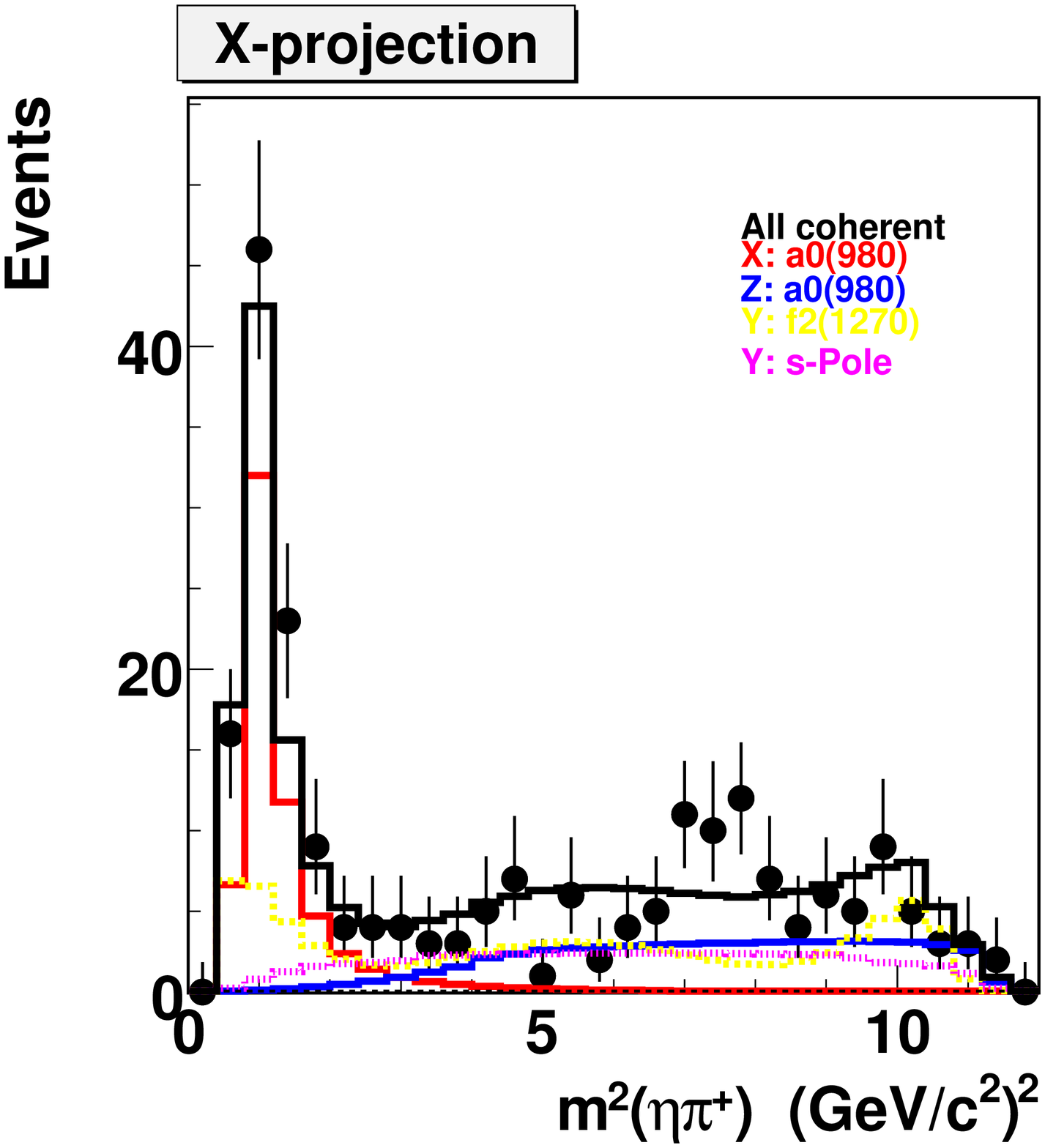}
\includegraphics*[width=41mm]{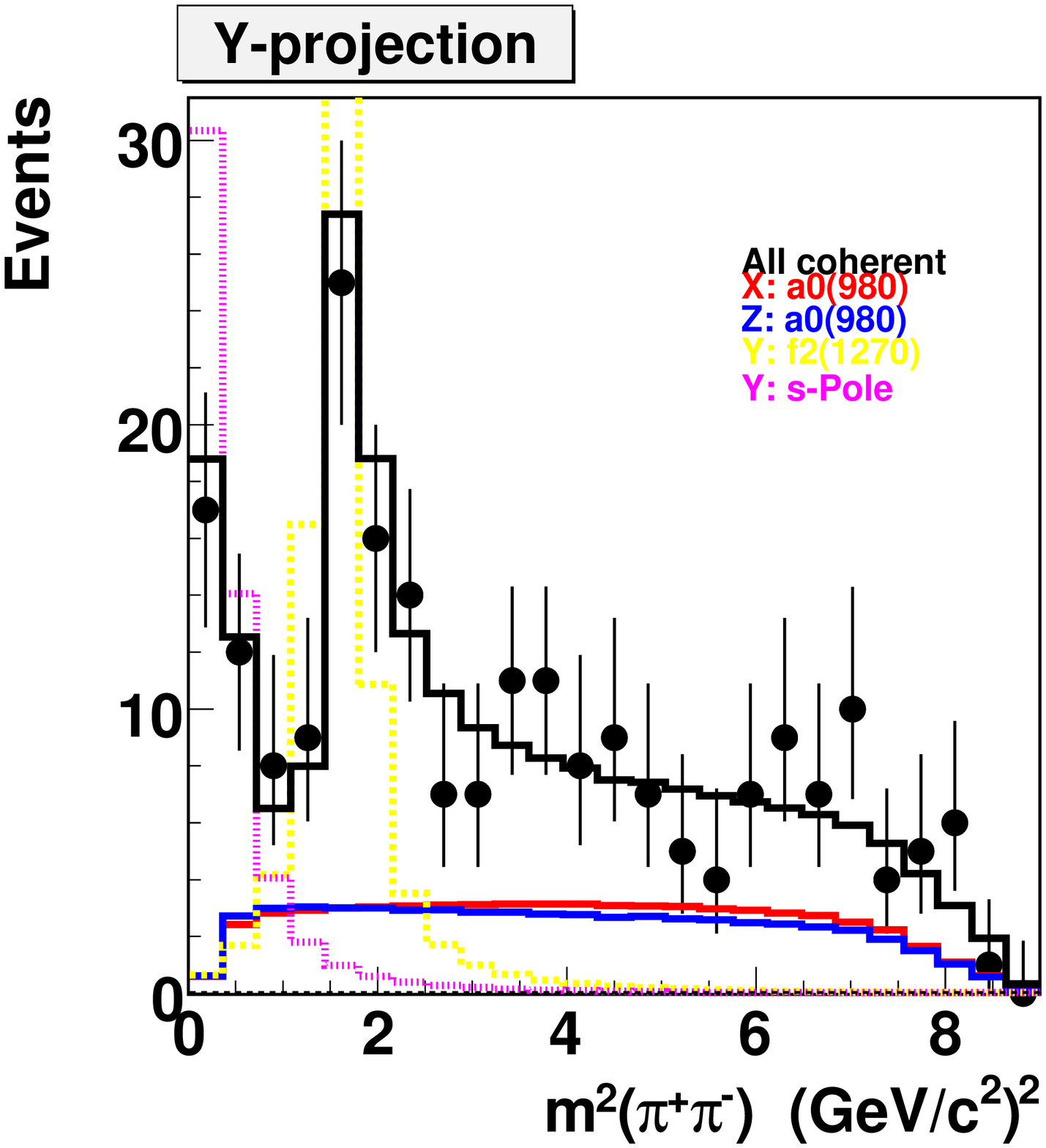}
\caption{Dalitz plot and projections on the two mass squared combinations
         for $\chi_{c1} \to \pi^+\pi^-\eta$.}
\label{fig:pipietadalitz}
\end{figure}

Fig.~\ref{fig:pipietadalitz} shows the Dalitz plot
and three projections for $\chi_{c1} \to \pi^+\pi^-\eta$.
There are clear contributions from $a_0(980)^\pm \pi^\mp$ and $f_2(1270) \eta$ intermediate states,
and significant accumulation of events at low $\pi^+\pi^-$ mass.
Isospin symmetry predicts that 
amplitudes and strong phases of both charge conjugated states should be equal.  
The overall amplitude normalization and one phase are arbitrary parameters and we set
$a_{a_0(980)^+} = a_{a_0(980)^-} = 1$,
$\phi_{a_0(980)^+} = \phi_{a_0(980)^-} =0$.
All other fit components are defined with respect to this choice for $a_0(980)$. 

Our initial fit to this mode includes only $a_0(980)^\pm \pi^\mp$ and $f_2(1270) \eta$
contributions, but has a low probability of describing the data, 0.13\%,
due to the accumulation of events at low $\pi^+\pi^-$ mass.
We find that only the $\sigma\eta$ describes well the low $\pi^+\pi^-$ mass spectrum,
and finally describe the Dalitz plot with 
$a_0(980)^\pm \pi^\mp$, $f_2(1270) \eta$, and $\sigma \eta$ contributions.
Table~\ref{tab:pipietadalitz} gives the preliminary results of this fit which has a 
probability to match the data of 58\%.
Systematic uncertainties shown in Table~\ref{tab:pipietadalitz}
were estimated from variations to nominal fit results in numerous cross-checks.
We consider variation of the efficiency shape, uncertainties in
resonance parameters, presence of additional resonances etc. 
We note that with higher statistics this mode
may offer one of the best measurements of the parameters of the $a_0(980)$.

\begin{table}[ph]
\tbl{ Preliminary fit results for $\chi_{c1} \to \pi^+\pi^-\eta$ Dalitz plot analysis.
      The uncertainties are statistical and systematic.}
{\begin{tabular}{l|l|l|c}
Contribution           & Amplitude               & Phase (${}^\circ$) & Fit Fraction (\%)  \\
\hline
$a_0(980)^\pm \pi^\mp$ & 1                       & 0                  & $56.2\pm3.6\pm1.4$ \\
$f_2(1270) \eta$       & $0.186\pm0.017\pm0.003$ & $-118\pm10\pm4$    & $35.1\pm2.9\pm1.8$ \\
$\sigma \eta$          & $0.68\pm0.07\pm0.05$    & $-85\pm18\pm15$    & $21.7\pm3.3\pm0.5$ \\
\end{tabular}\label{tab:pipietadalitz}}
\end{table}

\begin{figure}
\includegraphics*[width=41mm]{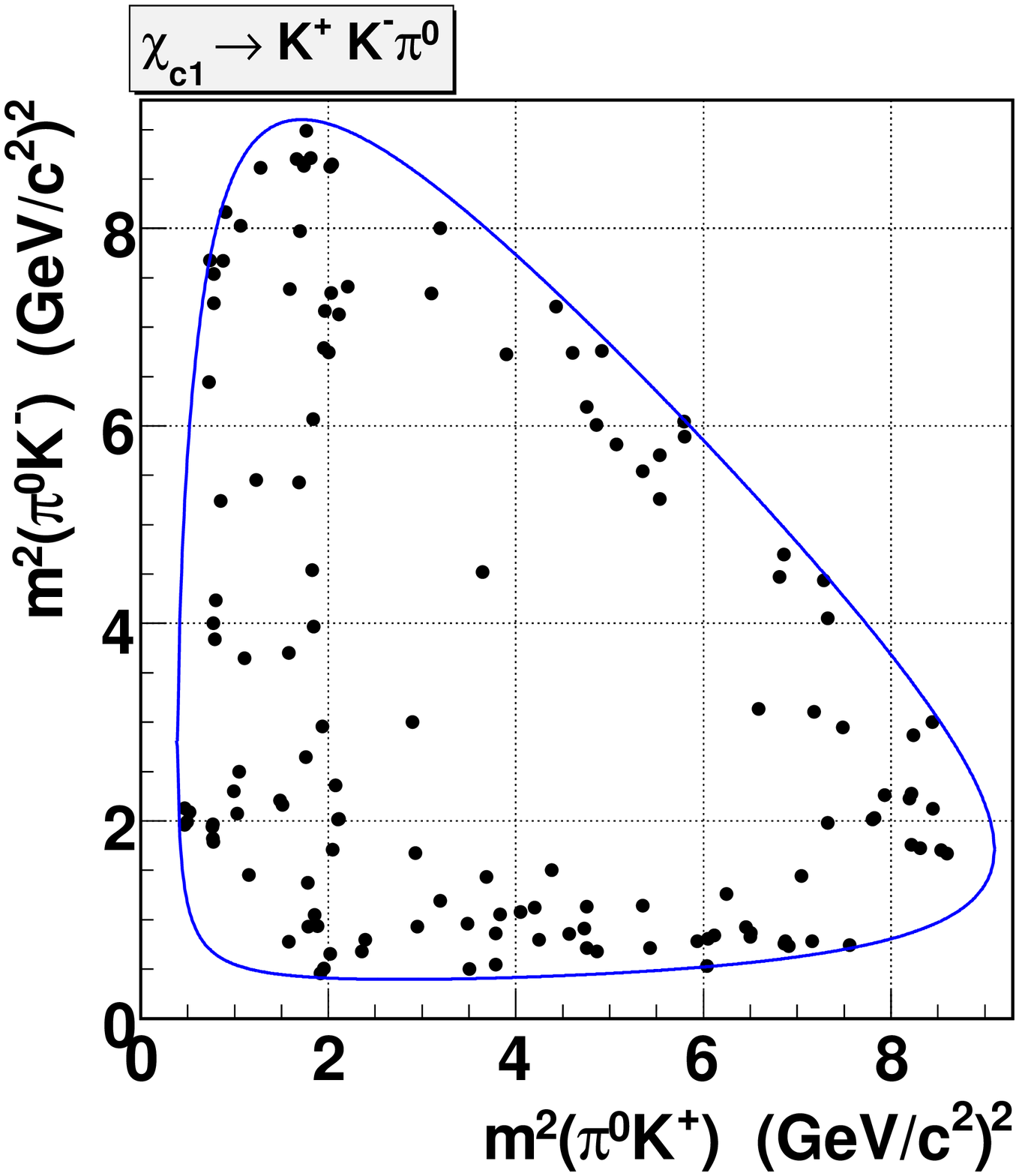}
\includegraphics*[width=41mm]{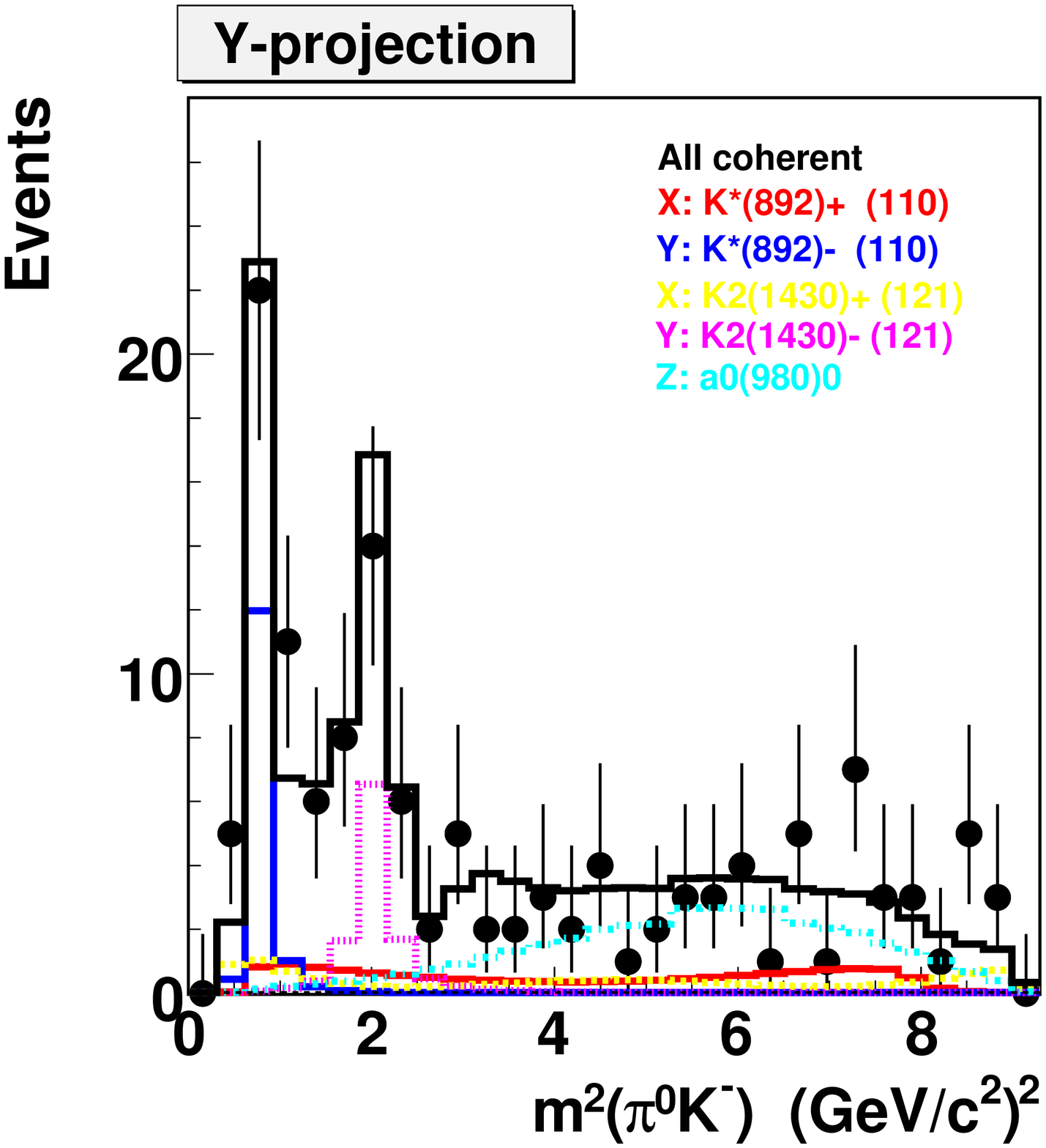}
\includegraphics*[width=41mm]{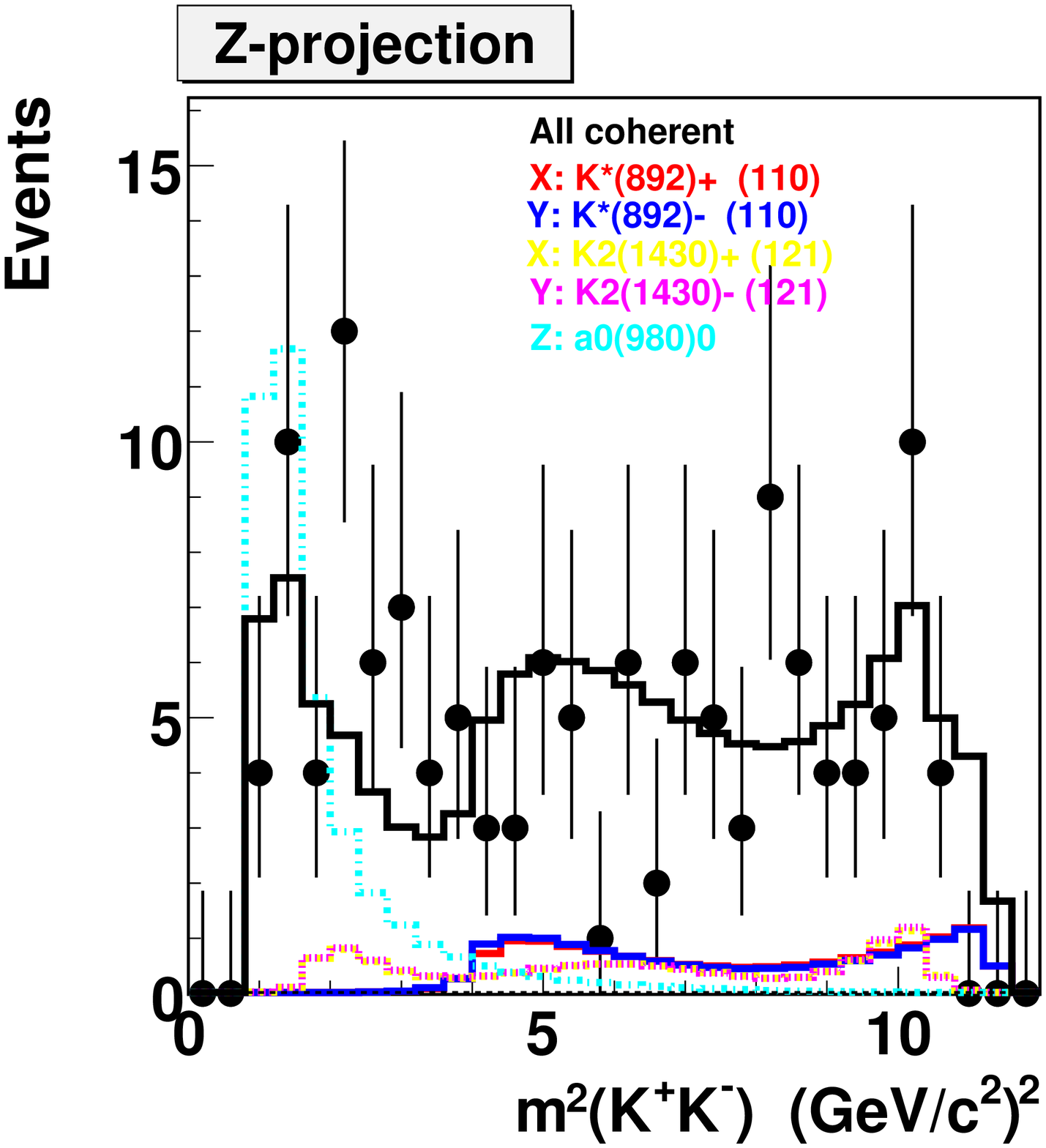}
\caption{Dalitz plot and projections on the two mass squared combinations
         for $\chi_{c1} \to K^+K^-\pi^0$.}
\label{fig:KKpi0dalitz}
\end{figure}
\begin{figure}
\includegraphics*[width=41mm]{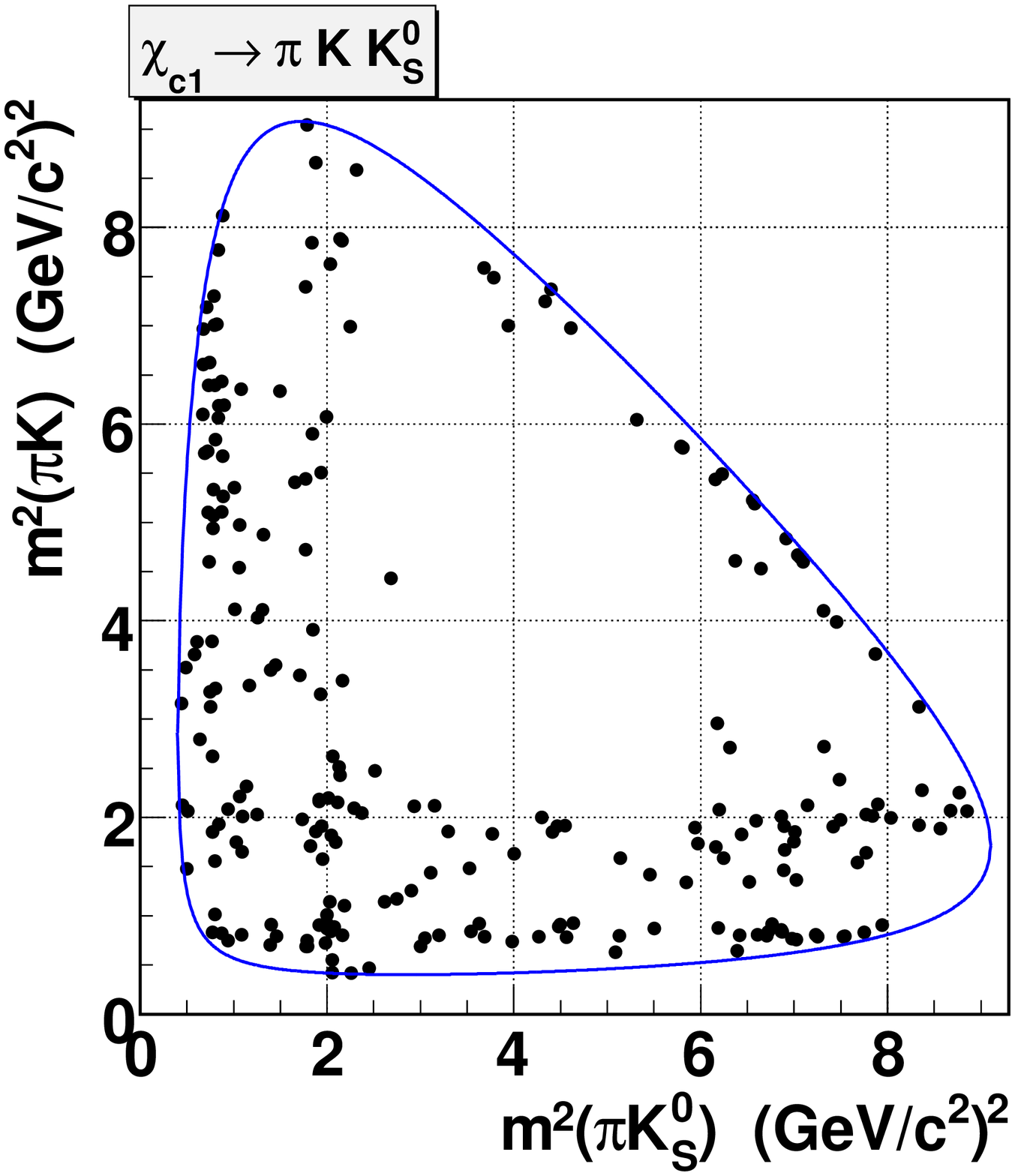}
\includegraphics*[width=41mm]{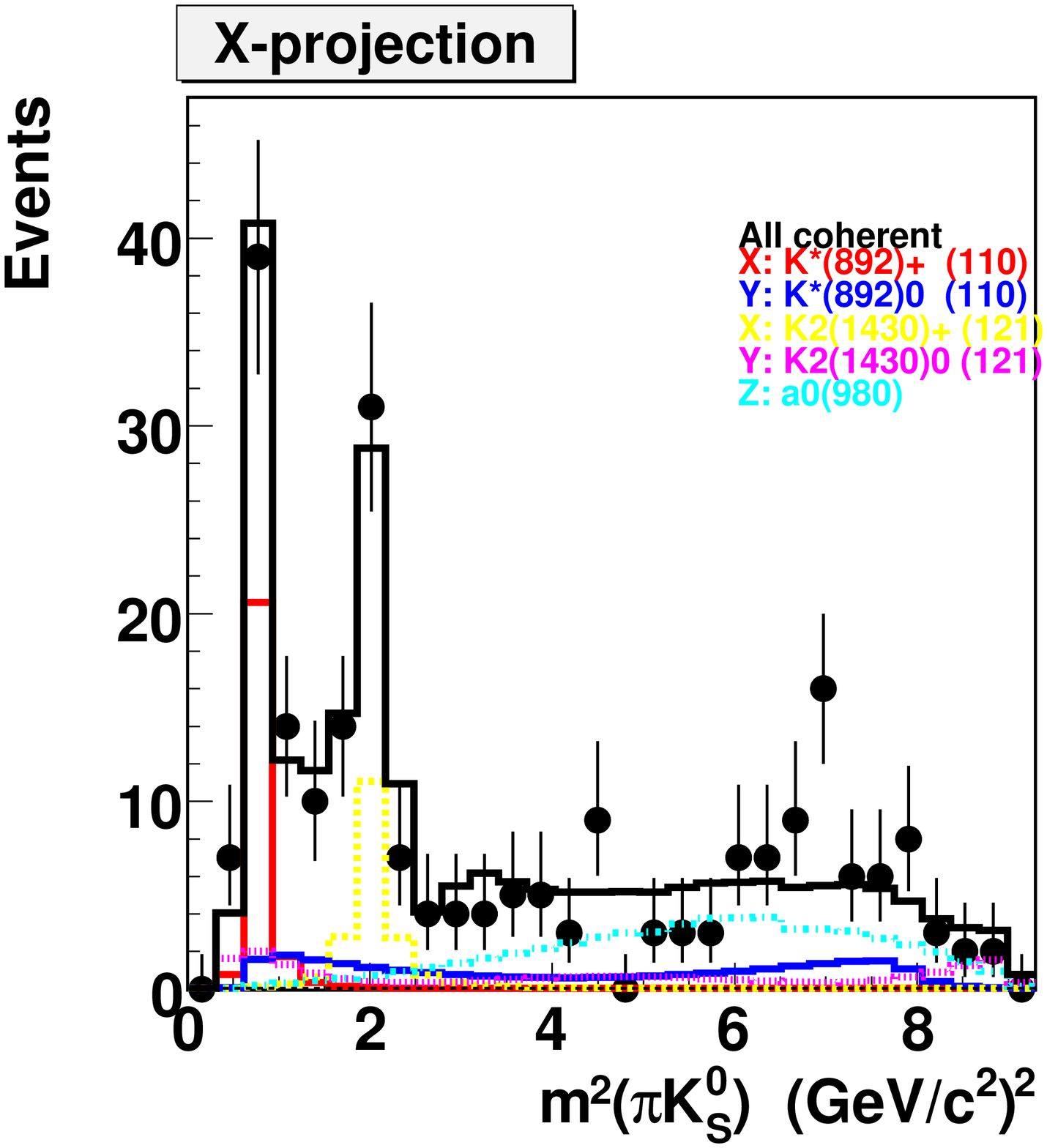}
\includegraphics*[width=41mm]{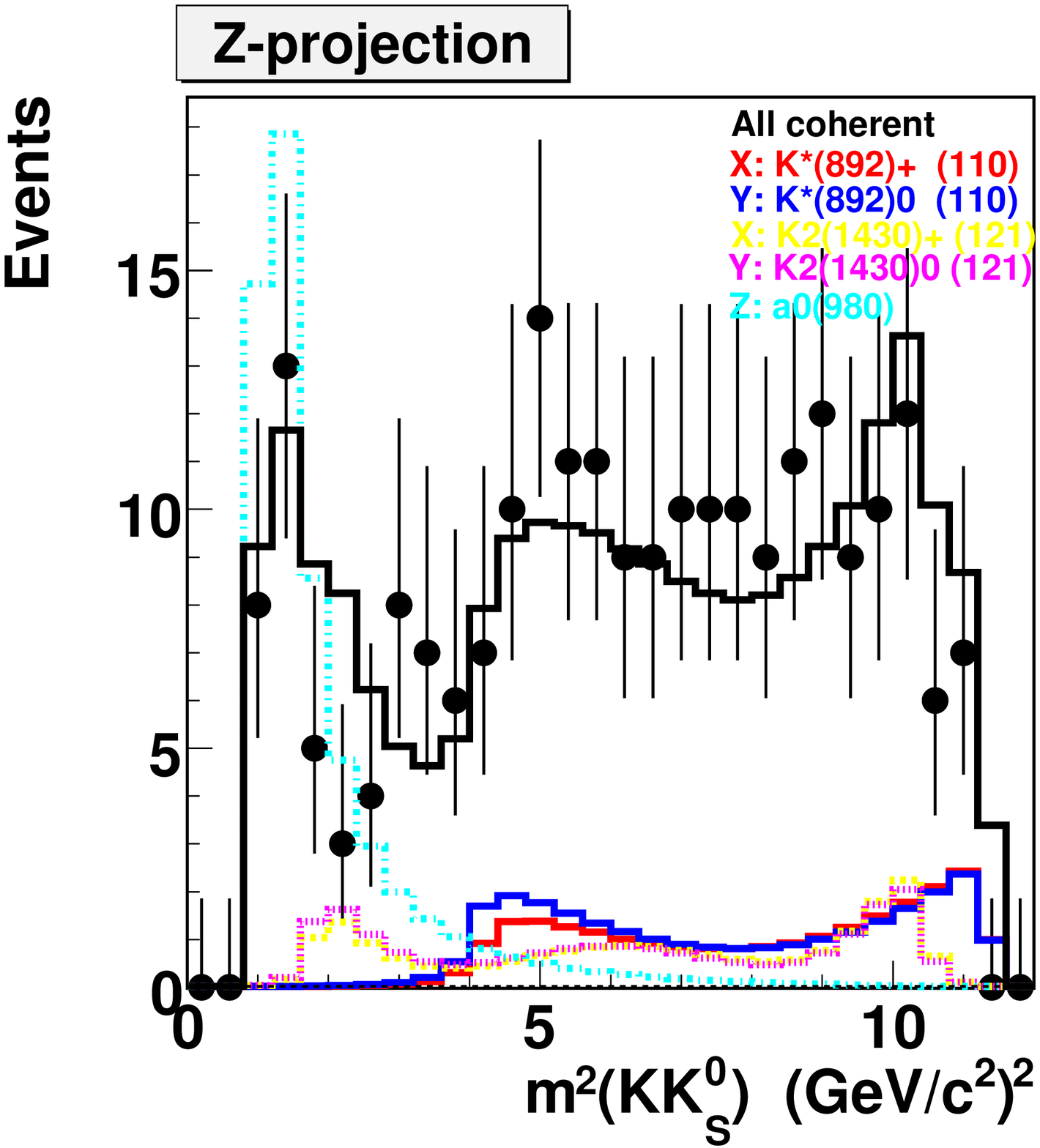}
\caption{Dalitz plot and projections on the two mass squared combinations
         for $\chi_{c1} \to K^0_SK^-\pi^+$.}
\label{fig:piKK0dalitz}
\end{figure}

The Dalitz plot for $\chi_{c1} \to K^+K^-\pi^0$ decay and
its projections are shown in Fig.~\ref{fig:KKpi0dalitz}, and
for $\chi_{c1} \to K^0_SK^-\pi^+$ in Fig.~\ref{fig:piKK0dalitz}.
We do a combined Dalitz plot analysis to these modes taking into account
isospin symmetry on amplitudes and phases:
$a_{K^{*+}} = a_{K^{*-}} = a_{K^{*0}} = a_{\overline{K^{*0}}} \equiv a_{K^{*}}$,
$\phi_{K^{*+}} = \phi_{K^{*-}} = \phi_{K^{*0}} = \phi_{\overline{K^{*0}}} \equiv \phi_{K^{*}}$,
$a_{a(980)^+} = a_{a(980)^-} = a_{a(980)^0} \equiv a_{a(980)}$, and
$\phi_{a(980)^+} = \phi_{a(980)^-} = \phi_{a(980)^0} \equiv \phi_{a(980)}$.
The common relative factor between two Dalitz plot amplitudes 
does not matter because of each Dalitz plot has an arbitrary normalization.
The overall amplitude normalization and one phase are arbitrary parameters and
we set
$a_{K^*(892)} = 1$ and
$\phi_{K^*(892)} = 0$.

The limited size of this sample, even in the combined Dalitz plot analysis,
and the many possible contributing resonances
leave us unable to draw clear conclusions.  Visual inspection shows clear
contributions from $K^*(892)^\pm K^\mp$, $K^*(892)^0K^0_S$,
$K^*(1430)^\pm K^\mp$, $K^*(1430)^0K^0_S$, 
$a_0(980)^0\pi^0$, and $a_0(980)^\pm\pi^\mp$.
It is not clear that the $K^*(1430)$ are $K^*_0$ or $K^*_2$, and many
other $K\pi$ and $KK$ resonances can possibly contribute.  
\begin{table}[ph]
\tbl{Preliminary results of the combined fits to the
     $\chi_{c1} \to K^+ K^- \pi^0$ and $\chi_{c1} \to K^0_S K \pi$ Dalitz plots.}
{\begin{tabular}{l|l|l|c}
Contribution           & Amplitude               & Phase (${}^\circ$) & Fit Fraction (\%)   \\
\hline
$K^*(892)K$            & 1                       & 0                  & $19.7\pm4.0\pm2.0$  \\
$K^*_2(1430)K$         & $0.50\pm0.09\pm0.12$    & $-2\pm13\pm6$      & $18.0\pm6.6\pm3.2$  \\
$K^*_0(1430)K$         & $5.3\pm1.0\pm0.1$       & $77\pm12\pm16$     & $36.0\pm12.8\pm3.0$ \\
$K^*(1680)K$           & $2.3\pm0.5\pm0.5$       & $-38\pm12\pm12$    & $11.2\pm5.4\pm2.7$  \\
$a_0(980)\pi$          & $10.8\pm1.2\pm1.2$      & $-112\pm12\pm3$    & $29.5\pm7.1\pm2.6$  \\
\end{tabular}\label{tab:KKpiDalitz}}
\end{table}
Our best fit
preliminary result is presented in Table~\ref{tab:KKpiDalitz} showing
statistical and systematic errors.  
This fit has a good 32\% probability of matching the data
and agrees with fits done to the separate Dalitz plots not taking advantage of
isospin symmetry.  
We also find that an alternative solution using the
same set of contributions fits the data acceptably, but less well than the 
displayed result.  


\section{Summary}
The CLEO collaboration possesses data samples
which can be used for analysis  
of $D$, $D_s$, $\chi_{cJ}$, $J/\psi$ meson decays.
Using roughly $10^6$ of $D^+D^-$ pairs 
produced in the $e^+e^- \to \psi(3770) \to D\overline{D}$ process
we perform a Dalitz plot analysis of the $D^+ \to \pi^-\pi^+\pi^+$ decay.
We confirm a dominant contribution from a low mass $\pi^+\pi^-$ S-wave amplitude
earlier observed by E791 and FOCUS experiments. 
Results of the $D^+ \to \pi^-\pi^+\pi^+$ Dalitz plot analysis 
are summarized in Table~\ref{tab:d3pidalitz}.

We have searched for and studied selected three body hadronic decays
of the $\chi_{c0}$, $\chi_{c1}$, $\chi_{c2}$ produced in radiative decays
of the $\psi(2S)$ in $e^+e^-$ collisions observed with the CLEO detector.  
The clean signal at least for one of $\chi_{cJ}$'s is found in eight final states:
$\pi^+\pi^-\eta $,
$\pi^+\pi^-\eta^\prime$,
$K^+K^-\pi^0$,
$K^0_S K \pi$,
$\eta K^+K^-$,
$\pi^0 p\bar{p}$,
$\eta p\bar{p}$, and
$\Lambda K^+\bar{p}$.
Our preliminary observations and branching fraction limits are summarized in 
Table~\ref{tab:Branching_fractions}.
In $\chi_{c1} \to \pi^+\pi^-\eta$ we have studied the resonant substructure using
a Dalitz plot analysis, and our preliminary results are summarized in Table~\ref{tab:pipietadalitz}.
Similarly in $\chi_{c1} \to KK\pi$ we clearly see contributions
from $K^*(892)K$ and $a_0(980)\pi$ at roughly the 20\% and 30\% level respectively, as shown in
Table~\ref{tab:KKpiDalitz}.


\section*{Acknowledgments}
We gratefully acknowledge the effort of the CESR staff
in providing us with excellent luminosity and running conditions.
D.~Cronin-Hennessy and A.~Ryd thank the A.P.~Sloan Foundation.
This work was supported by the National Science Foundation,
the U.S. Department of Energy, and
the Natural Sciences and Engineering Research Council of Canada.




\begin{thebibliography}{0}    
\bibitem{CPtagging} M.~Gronau, Y.~Grossman and  J.L.~Rosner, Phys.Lett. {\bf B 508} (2001) 37.
\bibitem{Dalitz} R.H.Dalitz, Phil. Mag. {\bf 44} (1953) 1068.
\bibitem{791} E.M.Aitala {\it et al.} (Fermilab E791 Collaboration), Phys. Rev. Lett. {\bf 86} (2001) 770.
\bibitem{focus} J.M.Link {\it et al.} (FOCUS Collaboration), Phys. Lett. {\bf B 585} (2004) 200.
\bibitem{CLEOdet} R.A. Briere {\it et al.} (CESR-c and CLEO-c Taskforces, CLEO-c Collaboration),
                  Cornell University, LEPP Report No. CLNS 01/1742 (2001) (unpublished).
\bibitem{cleodalitz} S. Kopp {\it et al.}, Phys. Rev. {\bf D 63}, 092001 (2001).
\bibitem{quarkoniumreview} N.~Brambilla {\it et al.}, CERN-2005-005 [hep-ph/0412158] is
                           a comprehensive recent review of heavy quarkonium physics.
\bibitem{pdg}  S. Eidelman {\it et al.}, Physics Letters {\bf B 592}, 1 (2004). 
\bibitem{chicbf} S.B. Athar {\it et al.}, (CLEO Collaboration), Phys. Rev. {\bf D 70}, 112002 (2004).
\bibitem{CBarrel_a0_980} Crystal Barrel Collaboration, Phys. Rev. {\bf D 57} (1998) 3860.
\bibitem{Oller_2005} J.A. Oller, Phys. Rev. {\bf D 71}, 054030 (2005).
\bibitem{Filippini-Fontana-Rotondi} V.~Filippini, A.~Fontana, and A.~Rotondi, Phys. Rev. {\bf D 51} (1995) 2247.

\end{thebibliography}
\end{document}